\documentstyle[twocolumn,epsfig,aps]{revtex}
\draft	

\begin{document} 
\twocolumn[\hsize\textwidth\columnwidth\hsize\csname
@twocolumnfalse\endcsname 

\title{Active Fusion and
Fission Processes on a Fluid Membrane}

\author{Madan Rao$^{1,2}$\cite{MAD} and
Sarasij R.\ C.$^{1,3}$\cite{SAR}
}

\address{
$^1$Raman Research Institute, C.V. Raman Avenue,
Sadashivanagar, Bangalore 560080, India\\
$^2$National Centre for Biological Sciences, UAS-GKVK Campus,
Bellary Road, Bangalore 560065, India\\
$^3$Institute of Mathematical Sciences, Taramani, Chennai 600113, India\\
}


\date{\today}

\maketitle

\begin{abstract}

We investigate the steady states and dynamical instabilities resulting from
``particles'' depositing on (fusion) and pinching off (fission)  a fluid
membrane. These particles could be either small lipid vesicles or isolated
proteins. In the stable case, such fusion/fission events suppress long
wavelength fluctuations of the membrane. In the unstable case, the membrane
shoots out long tubular structures reminiscent of endosomal compartments or
folded structures as in internal membranes like the endoplasmic reticulum or
Golgi. 


\end{abstract}

]
\vskip1.0in

The plasma membrane and internal membranes of eukaryotic cells such as
endosomal compartments, Golgi and endoplasmic reticulum (ER) are
subject to random deposition and evaporation of `particles'\cite{CELL}.
These particles could be isolated lipids or membrane proteins but most often
are small vesicles carrying cargo. For instance, in the endocytic
pathway, small vesicles (diameter $\approx 50$nm), pinch off from the plasma
membrane and fuse with 
the sorting endosome. 
The sorting endosome is seen to possess
long, thin, tubular protrusions that sporadically break away from
it\cite{REVIEW}. Likewise the ER $\longleftrightarrow$ Golgi
$\longleftrightarrow$ Plasma Membrane system\cite{SNARE} witnesses a major
trafficking of protein carrying vesicles. The Golgi and the ER are
highly ramified membranal structures, often possessing long thin tubular
protusions\cite{CELL}.

The fusion/fission events are `active' processes in that
they are triggered by signalling proteins which recruit an elaborate
array of proteins as part of the fusion/fission
machinery\cite{CELL,SNARE}. These proteins, such as coat proteins,
SNARES and rabs are functionally activated by energy provided by the
consumption of ATP/GTP. Though at first sight the phenomenon appears
alarmingly complex, we argue below that it gives rise to
precise questions of a generic nature that do not require a knowledge
of specific biomolecular details, and which may be addressed within the
framework of
nonequilibrium statistical physics.

In this Letter we ask\,: Starting with an almost planar
fluid membrane, what is its eventual shape when subject to such active
fusion/fission events ? We will find that 
the final fate of the membrane is
(a) a steady state shape whose long wavelength fluctuations are suppressed
by such events or (b) a dynamical instability giving rise to either long
tubular structures or folded structures which grow rapidly. This rapid
growth would be halted by other processes (not included in our analysis)
which might lead to an eventual pinch off from the membrane.

We take a coarse-grained approach 
in which the equilibrium shapes of a fluid membrane (in
the absence of fusion/fission),
parametrized by its 3-dimensional conformation ${\bf R}(u_1,u_2)$ as
a function of local coordinates $(u_1,u_2)$, are governed by the
local curvature\cite{CURV} and the areal density $\tilde \psi$ of
`particles' on it, through the effective free-energy functional
\begin{equation}
F_{el} = \frac{1}{2} \int d^{2}u {\sqrt {g}} \left[ 
\sigma +
\kappa_c H^2 +  2 c_0 {\tilde \psi} H + \chi {\tilde \psi}^2 \right]\,.
\label{eq:ELASTIC}
\end{equation}

The local mean curvature $H = 1/2\, g_{ab}{\hat{\bf N}}\cdot\,\partial_a
\partial_b {\bf R}$, where ${\hat{\bf N}}$ is the local unit (outward)  
normal and $g_{ab}$ is the induced metric ($g\equiv$\,det\,$g_{ab}$). The
density field ${\tilde \psi}$ has the following interpretation --- when
the particles are tiny lipid vesicles, ${\tilde \psi} = {\tilde \rho} -
{\tilde \rho}_0$ is the excess areal lipid density (${\tilde \rho}$ is the
local areal lipid density and ${\tilde \rho}_0$ is its equilbrium value)
and $c_0=0$\,; when the particles are isolated membrane proteins or
structurally asymmetric transmembrane proteins, ${\tilde \psi}$ is the
areal protein density and the spontaneous curvature modulus $c_0\neq0$,
reflecting the change in the local elastic properties of the membrane due
to the embedded protein.

In this paper we will work entirely in the Monge representation where
the membrane shape is represented by a height field $h(x,y)$ and all
physical quantities are computed in the horizontal ${\bf
x}\equiv(x,y)$ plane. In this representation $g={1+(\nabla h)^2}$ and
$H = \nabla^2 h /g^{3/2}$. The areal densities projected onto the
${\bf x}$ plane are obtained from $\rho = {\tilde \rho}{\sqrt {g}}$. 
In our notation quantities with and without the tilde will denote
intrinsic and projected densities respectively.
In this Monge representation, thermal equilibrium height correlations are
given by $ \langle \vert h_{q} \vert^2 \rangle \propto T/\left(\sigma q^2
+ \kappa_{R} q^4 \right)$, where $\kappa_{R} = \kappa_{c} -
\left(\kappa_{c} c_0\right)^2/\chi$.

We now start the fusion/fission processes which drive the membrane
out of equilibrium, giving rise to a dynamical excess density field
$\psi({\bf x},t)$ and a dynamical shape $h({\bf x},t)$. For
convenience we shall use the {\it picture} of fusion/fission of
small vesicles in the rest of the paper, although our
mathematical treatment
includes the deposition and `evaporation' of
isolated proteins as well.
Our subsequent analysis is valid over length
scales much larger than the typical size $\ell$ of the `particles'
and time scales much longer than the duration of individual
fusion(fission)-events $\tau$. 
Thus we shall ignore 
the complicated microscopic processes at work during the event. 
As in \cite{PUMPS}, we try to extract the most robust 
features through general symmetry arguments.
To facilitate analysis, we can regard a local
region of the membrane as being in one of three states: (i) an
undistorted membrane; (ii) a membrane with a `bump'; (iii) an
undistorted membrane with a proximate vesicle.  The balance between
(i) and (ii) is governed largely by elastic and equilibrium
thermodynamic forces whereas that between (ii) and (iii) is
determined mainly by active processes \cite{FOOT1} (Fig.\ 1).
\begin{figure} 
\centerline{\epsfig{figure=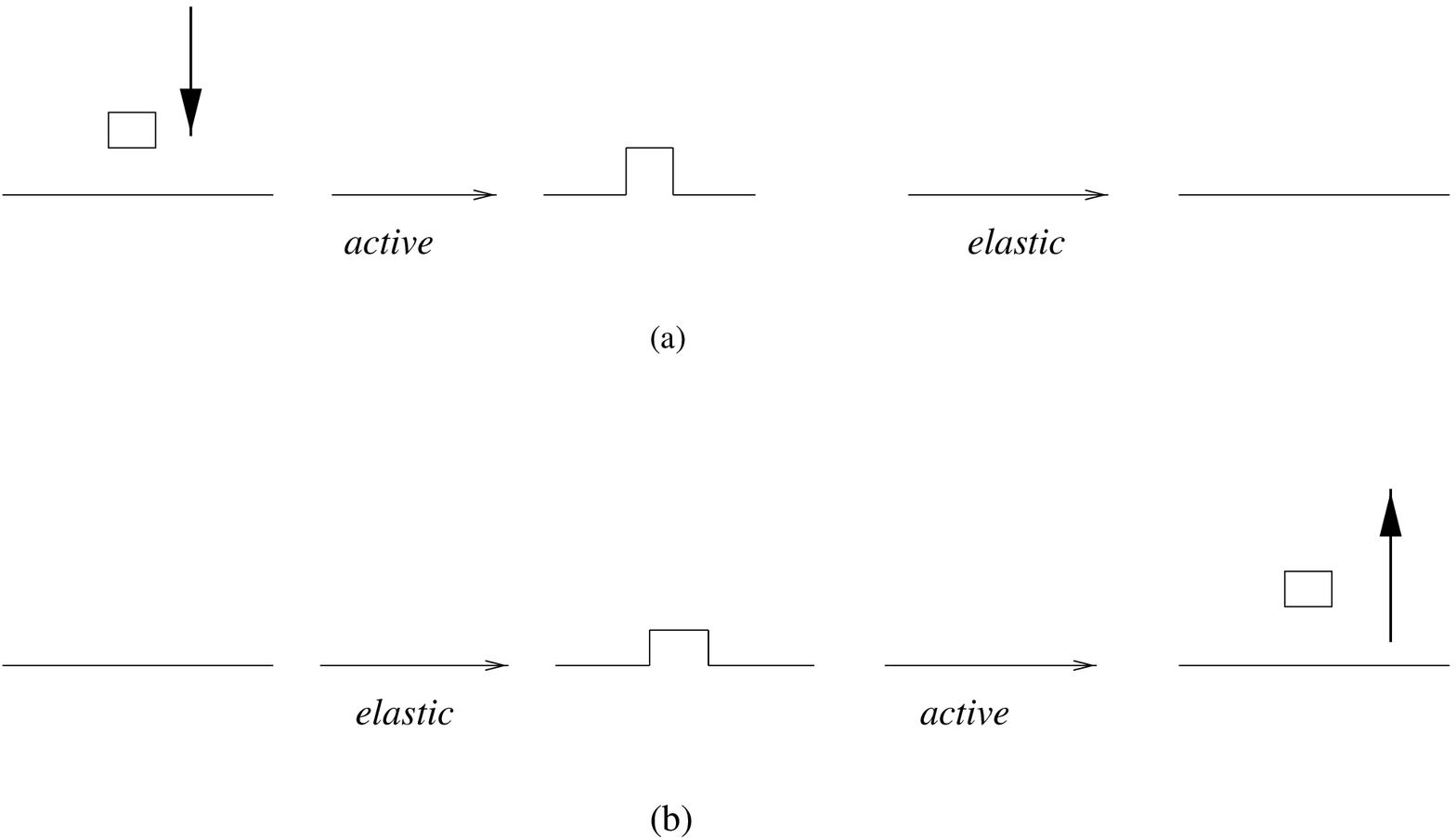,width=6.0cm,height=5.0cm}}
\end{figure} 
FIG. 1. The ``active'' and ``elastic'' stages of (a) fusion  and
(b) fission.  
\\ 

Define an `fusion-event' $\epsilon_{+}({\bf x},t)$ and an
`fission-event' $\epsilon_{-}({\bf x},t)$ as random numbers taking
values $0$ or $1$, with a nonzero mean $\langle \epsilon_{\pm}\rangle
= e_{\pm}$ and distributed identically and independently at different
${\bf x}$ and $t$.  
Every event gives rise to a change in the local height and density\,;
we would therefore like to calculate the rate of such fusion/fission
events (number of events/area/time).  We argue as follows\,:  Since
the fusion/fission-events happen only on the cytosolic side in the
case of internal membranes, and in an in-out asymmetric manner in the
case of the plasma membrane, a local observer on the membrane can
tell up from down in the Monge description. Thus all dynamical
effects should reflect this up/down membrane asymmetry. This implies
that the rate of fusion(fission) events, $r^{+}(r^{-})$, should be
sensitive to the local mean curvature 
and hence to
lowest order can be written as
\begin{eqnarray}
r^{+}_n  &  =  &  \lambda_1 + \lambda_2 H + \lambda_3 {\tilde \rho} +
\lambda_4  {\tilde \rho} H + \ldots
\\
r^{-}_n  &  =  &  \mu_1 {\tilde \rho} + \mu_2 {\tilde \rho} H + \ldots
\label{eq:RATES}
\end{eqnarray}
where the coefficients $\lambda_1, \, \mu_1 > 0$ by virtue of the definition
of the fusion/fission processes.  Note that symmetry arguments say nothing
about the values of the other parameters, whose magnitudes and signs are
determined by the microphysics of fusion/fission.

To determine the net normal velocity, we multiply the above rates by
a typical velocity $\ell/\tau$ and include a permeative contribution
proportional to the force $f_{el}\equiv \delta F_{el}/\delta {\bf
R}$,
\begin{equation}
v_n = (\ell/\tau)\, \left[ \epsilon_{+}(x,t)\,r^{+}_n -
\epsilon_{-}(x,t)\,r^{-}_n
\right]
    - \Gamma f_{el}\,,
\label{eq:NORMAL}
\end{equation}
where $\Gamma$ is a mobility.
Similarly the changes in the density
due to the fusion/fission events are proportional to the ``particle
currents'',
\begin{eqnarray}
j_{+}    &   =   &  \left(A_{+}/\tau\right) \, \epsilon_{+}({\bf x},t)\,r^{+}_n 
\label{eq:CURRENT+}\\  
j_{-}   &   =   &  \left(A_{-}/\tau\right) \, \epsilon_{-}({\bf x},t)\,r^{-}_n
\label{eq:CURRENT-}
\end{eqnarray}
where $A_{\pm}$ are dimensionless, positive constants. It will turn out
that we may consistently take $A_{+}=A_{-}$ (for lipid vesicles,
$A_{+}=\ell^2/a$, the amount of lipid material brought during a typical
fusion event, $a$ being the area per lipid).


In the Monge representation, we decompose the normal velocity along
the vertical $z$ axis, $v_z = v_n ({\hat {\bf N}} \cdot {\hat {\bf
z}})$, and the horizontal ${\bf x}$ plane, $v^{\alpha}_{\perp} = v_n
({\hat {\bf N}} \cdot {\hat {\bf x}}^{\alpha})$, where $\alpha = 1,2$
labels the basis vectors in the x-y plane\cite{CAI}. 
Using the definitions (2) to (6) this yields the equations of motion,
\begin{eqnarray}
\partial_t h    &   =   &  \frac{\ell}{\tau}\, \left[ \epsilon_{+} \left(
\lambda_1
+
\lambda_2 \nabla^2 h + \lambda_3 \rho + \lambda_4 \rho \nabla^2 h \right)
\right.  \nonumber \\
                &       & \left. - \epsilon_{-} \left( \mu_1 \rho +
\mu_2 \rho
                      \nabla^2 h \right)\right] \nonumber \\
                &       &  - \Gamma 
\left[ -\sigma \nabla^2 h + \kappa_c \nabla^4 h - \kappa_c c_0 \nabla^2
\rho 
\right]
+ v_{h} + f_h
\label{eq:HEIGHT}
\end{eqnarray}
for the height field and
\begin{equation}
\frac{\partial \rho}{\partial t}   =    - \,\nabla \cdot (\rho {\bf
                       v}_{\perp}) + \nabla \cdot \left(D \nabla 
                        \frac{\delta
                        F_{el}}{\delta
                       \rho}\right) + j_{+} - j_{-} + \nabla \cdot {\bf
                        f}_{\rho}\,,
\label{eq:DENSITY}
\end{equation}
for the density field\cite{FOOT2}.
In Eq.\ (\ref{eq:HEIGHT}), $v_{h} = - \int \frac{d^2 q}{(2\pi)^2}
e^{i {\bf q}\cdot{\bf x}} \frac{1}{4\eta \vert q\vert} \frac{\delta
F_{el}}{\delta h_q(t)}$, is the vertical component of the viscous
flow ($\eta$ is the solvent viscosity) induced at the membrane surface by
the membrane elastic stresses \cite{CAI}.
We have ignored other sources of
dissipation coming from intramembrane viscosity and bilayer
friction\cite{SEIFERT}.
The equations of motion also contain thermal noise sources having
zero mean\,; in the Monge representation, the variance
$\langle \vert f_h({\bf q},t)\vert^2\rangle = 2T \left[
\Gamma + \left(4 \eta \vert q\vert\right)^{-1} \right]$, while the 
variance of the vector noise ${\bf
f}_{\rho}$ is $2 T D$. 
Note the source and sink terms $j_{\pm}$ in Eq.\ (\ref{eq:DENSITY}) do not
conserve particle material.


Let us first check if Eqs.\
(\ref{eq:HEIGHT},\ref{eq:DENSITY}) admit
steady state solutions when we replace the noise terms by their
means. We
shall drop the hydrodynamic force for
simplicity. Our subsequent analysis will be for a tensionless
($\sigma =0$) membrane with no spontaneous curvature ($c_0 = 0$). 
We find
{\it spatially homogeneous} steady states for which $j_{+}=j_{-}$, which
implies that in the steady state the density attains a uniform value
given by
\begin{equation}
\rho_{\infty} = - \frac{e_{+}\lambda_1}{e_{+} \lambda_3 - e_{-}\mu_1}\,,
\label{eq:STEADY}
\end{equation} 
while the steady state height is a constant (which we take to
be $0$). Since $\lambda_1 > 0$, the
above steady state holds when $B\equiv e_{+} \lambda_3 - e_{-} \mu_1 < 0$. 
Is this steady state linearly stable to 
perturbations in $h$ and $\rho = \rho_{\infty} +
\phi$ ?  In momentum space, the linearised perturbations are given by,
\begin{equation}
\frac{d}{dt} \left( \begin{array}{c}
                          h_q  \\  \phi_q
                      \end{array} \right) = 
            \left( \begin{array}{lr}
 - \frac{\ell}{\tau}\,A q^2 - \kappa_c \Gamma q^4  &  \frac{\ell}{\tau}\,B
\\
 - \frac{\ell^2}{a \tau} A q^2           &  \frac{\ell^2}{a \tau}\, B - D
\chi q^2
                         \end{array} \right)
             \left( \begin{array}{c}
                          h_q       \\
                         \phi_q 
                     \end{array} \right)
\label{eq:MODES}
\end{equation}
where $A = e_{+} \lambda_2 + \rho_{\infty} (e_{+} \lambda_4 -
e_{-} \mu_2)$. The
eigenvalues of the dynamical matrix for small $q$ are
\begin{equation}
\lambda_{1,2} = \left\{ \begin{array}{l}
                 \frac{\ell^2}{a \tau}\,B - {q^2} \left[ D \chi +
\frac{\ell}{\tau}\,A
\right] \\
                 - q^4 \left[ \kappa_c \Gamma - \frac{\ell}{(\ell^2/a
\tau)}\,A\,D\chi/B
\right]
                       \end{array}
                 \right. 
\label{eq:EIGEN}
\end{equation}
The nonconservative $\phi$ ensures that one of the modes is always
`fast' {\it i.e.}, has a nonzero relaxation rate at sufficiently long
wavelength.
If $\ell\,A/\tau >- \min\,[\,D\chi\,,\,\kappa_c \Gamma \,(\ell^2/a
\tau)\,\vert
B\vert/D\chi] = - A^*$,
then both modes decay and 
the homogeneous flat steady
state is stable. We calculate the equal-time height fluctuations about
this steady
state, by first restoring the thermal 
and non-thermal noises $\epsilon_{\pm} = e_{\pm} + \delta \epsilon_{\pm}$,
in Eq.\ (\ref{eq:MODES}), fourier
transforming
with respect to ${\bf x}$ and $t$, and then integrating the correlator
with
respect to frequency $\omega$ to obtain,
\begin{equation}
\langle h_{q}(t) h_{-q}(t)\rangle = \frac{E}{
q^2}+\frac{T}{\frac{\ell A}{\Gamma}\,q^2
+
\kappa_c \,q^4}
\label{eq:CORRELATOR}
\end{equation}
for small $q$. The first term is a {\it purely nonequilibrium}
contribution
to the height fluctuations with a $q^2$ dispersion and an
amplitude $E$ related to the variance of 
$\delta \epsilon_{\pm}$ and the activity.
The second term is a
modification of the usual
thermal correlator --- the fission/fusion 
activity {\em suppresses} height fluctuations at small $q$ 
through an activity-induced tension (see \cite{PUMPS})  
proportional to $A$. On the other hand, density
correlations are short ranged with a correlation length $\xi =
B^{-1/2}$. 

If $A < -A^*$, then either one or both modes are unstable.
The unstable massless mode grows as $q^4$ for small $q$. 
Physically this instability is a consequence of the curvature
dependent rate of fusion\,: an initial outward bump
($\nabla^2 h < 0$)  will promote a faster rate of fusion, adding
on more $h$. 

Going beyond linear stability analysis in the case $A < - A^*$ 
requires a numerical simulation of Eqs.\
(\ref{eq:HEIGHT},\ref{eq:DENSITY}). It is convenient to work in
dimensionless variables, where $x \to x/{\sqrt{\kappa_c/\chi}}$, $t \to
t/(\kappa_c/D\chi)$, $h \to h/\ell$ and $\rho \to \rho/(\ell/a^2)$. Using
an
Euler discretisation on a square grid of size $N^2 = 100^2$, we convert
the differential equations to a coupled map. The spatial and temporal
increments have been chosen to satisfy $\triangle t <
\Gamma^{-1}\,(\triangle x)^4$ and $\triangle t < (\triangle x)^2$ 
to ensure numerical stability. For most of
our analysis we have fixed $\triangle x = 0.1$ and $\triangle t =
10^{-4}$.  We have taken open boundary conditions and have used a large
enough system to ensure that no finite size effects corrupt our
interpretation. We report numerical results only for the linearly unstable
case.

We start with the homogeneous steady state, $h({\bf
x},0)=0$ and 
$\rho=\rho_{\infty}$, and set the non-thermal noises to their
average values, $\epsilon_{\pm} = e_{\pm}$. Introduce a small bump, a
paraboloid of height $L_0$ and width $w_0$ smoothly connected to the
rest of the membrane. The peak height $L(t)$ and the width $w(t)$
grow initially, the width then saturates at $q_{*}^{-1}\sim \vert
A\vert^{-1/2}$, the length scale corresponding to the fastest growing
mode determined from the linear analysis. The height now grows
exponentially as $L(t) \sim \exp(A^2 t)$ (Fig.\ 2). Beyond this
linear regime, whose time scale is set by $A$, the nonlinear term
$\nu \rho \nabla^2 h$ (where $\nu \equiv e_{+} \lambda_4 - e_{-}
\mu_2$) starts becoming comparable.  The sign of $\nu$ dictates the
subsequent evolution\cite{KPZ}.  If $\nu < 0$, the nonlinear term {\it
accelerates} the growth, producing a long thin tubule (Fig.\ 2).
Since the local curvature of the membrane becomes positive in the
region where the main peak joins the rest of the membrane, this gives
rise to side branches which dig into the membrane at a much slower
rate (Fig.\ 2). The resulting morphology is a long tubule which grows
very fast. This fast growth may eventually be halted by molecular
processes not contained in our coarse-grained model, leading to a
fissioning of the tubule from the parent membrane. 
\vskip -2.0cm
\begin{figure} 
\centerline{\epsfig{figure=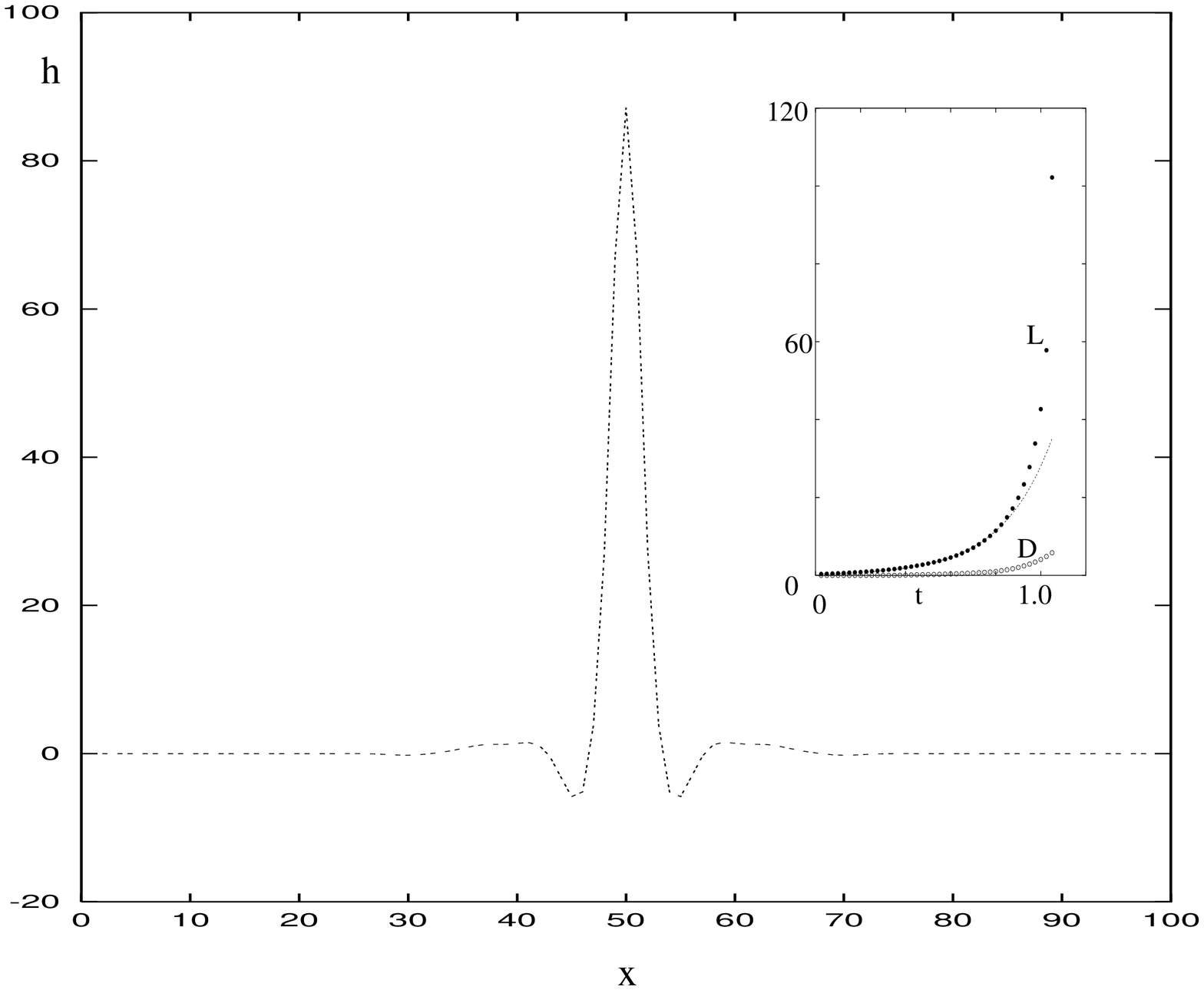,width=6.0cm,height=8.0cm}}
\end{figure}
\vskip 0.5cm
FIG. 2. Tubule configuration (obtained by rotating the profile about the
vertical axis) at $t=1.1$ when $\nu <0$.
Inset shows the peak
height $L$ and the depth of the side branch $D$ (dashed line is the
initial exponential growth of $L$). The parameters $\nu = -0.012$ and
$A=-0.16$.\\

When $\nu >0$, the exponential rise of the peak height is cut off by a
{\it
decrease} in the rate $j \equiv j_{+} - j_{-}$. At the same time the
density at the peak diffuses out into the flanks in such a way as to
compensate for the increase coming from the addition of `particles'. This
leads to a saturation of the density at the peak and asymptotically
approaches $\rho_s = - e_{+}\lambda_2/\nu$. The depth of the flanks
increases exponentially fast till it is comparable with the slowly
increasing peak height $L$. The resulting structure is a folded
configuration as in Fig.\ 3.

So far we had set the non-thermal noises $\epsilon_{\pm}$ to their
average values\,; we had therefore needed an initial bump on the membrane
to seed the growth of tubules. The situation is unchanged if instead we
let $\epsilon_{\pm}({\bf x},t) = e_{\pm} + \delta \epsilon_{\pm}({\bf
x},t)$, where the fluctuations are uncorrelated in space and time.  In
practice, this initial localised perturbation about the steady state may
be achieved by either phase separation of different lipid species
constituting the membrane \cite{REVIEW} or noise. If we however
set $\epsilon_{\pm}({\bf x},t) = \epsilon_{\pm}(t) \delta({\bf x})$,
where $\epsilon_{\pm}(t)$ is stochastic,
then the membrane {\it spontaneously creates a bump at the
fusion site}, growing exponentially fast.
\vskip -3.0cm
\begin{figure} 
\centerline{\epsfig{figure=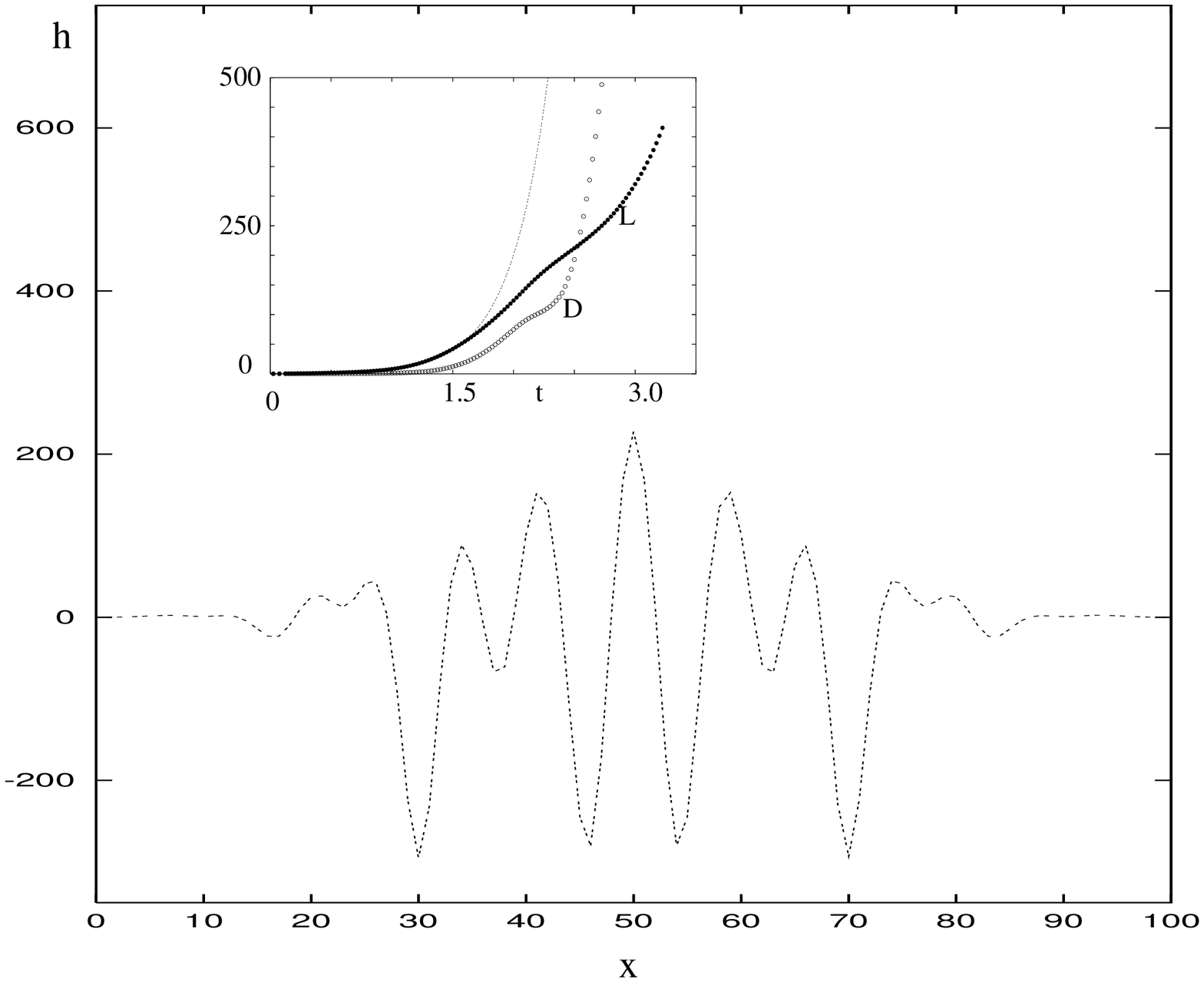,width=6.0cm,height=8.0cm}}
\end{figure}
\vskip 0.5cm
FIG. 3. Folded configuration (obtained on rotation about the vertical) at
$t=2.6$ when $\nu >0$. Inset
shows the height of the central peak $L$ and the
largest depth of the folds $D$ (dashed line is the initial
exponential growth of $L$). The parameters $\nu = 0.012$ and
$A=-0.14$.\\

The tubular extensions produced when $\nu <0$ look similar to the
structures exhibited by the sorting endosome associated with the
endocytic pathway of transferrin and LDL proteins. The sorting
endosome is seen to possess several tubular structures of width
$\approx 50$nm which grow to a length of about a micron before
peeling off from the parent structure\cite{REVIEW}. On the other
hand, the folded structures produced when $\nu>0$ are most
dramatically exhibited in a closed membrane, which may be studied
using a dynamical triangulation monte carlo modified to add and
subtract particles \cite{DTMC}. 

Such tubulation or folded morphologies could also be generated if the
`particles' depositing on and `evaporating' from the membrane are
isolated proteins. The rates of these events may be altered by the
action of some drug, such as Brefeldin-A. Our analysis suggests that
such deposition-evaporation of proteins from a membrane could result
in tubulation\cite{JITU}.


A second look at Eqs.\ (\ref{eq:HEIGHT},\ref{eq:DENSITY}) allows us
to make some definite predictions. We see that the surface tension,
which is the coefficient of the $\nabla^2 h$ term, gets renormalized
by an `activity induced' factor, ${\tilde \sigma} = \sigma +
\ell\,A/\Gamma$.  This renormalization is also reflected in the form of
the correlators, Eq.\ (\ref{eq:CORRELATOR}). Alternatively, {\it
applying a tension $\sigma$ to the membrane will affect the rate of
fusion/fission events} \cite{SHEETZ}. 

The activity induced tension is in general inhomogeneous since the
`activity' $A$ may be space dependent. This gives rise to gradients
in the activity induced surface tension, driving flow of membrane
constituents towards regions of lower tension. Ignoring bilayer
friction, the velocity of this Marangoni flow may be written as ${\bf
v}_{\perp} = - \frac{\xi^2}{\mu} \nabla {\tilde \sigma}$, where $\mu$
is the in-plane shear viscosity and $\xi$ is a length scale
associated with the `particle' size. The direction of flow is such
that the membrane constituents will move towards the tubular
extensions produced as a result of the activity. 

In future we would like to study the effects of stochastic fusion/fission
of `particles' on a closed fluid membrane\cite{DTMC}, and to explore the
complete nonequilibrium steady state diagram by solving the
stochastic equations (\ref{eq:HEIGHT},\ref{eq:DENSITY}).

We thank P.\ B.\ S.\ Kumar and especially S.\ Mayor and S. Ramaswamy
for several illuminating discussions and critical readings of the
manuscript.  We thank J.\ Prost for informing us of a short note on the
same subject\cite{BRUSH}. MR thanks the Department of Science and
Technology, India for a Swarnajayanti Fellowship.


\begin{references}
\bibitem{FOOT1} We should distinguish this active fusion/fission process
from the deposition/evaporation of particles onto/from an interface
separating two bulk coexisting phases.
\bibitem{FOOT2} In principle, the equations of motion in the case when
the `particles' are isolated proteins should include the height $h$, the
nonconserved protein density $\rho$ and the conserved lipid density
$c$. The lipid density however does not couple to the other two fields.
\bibitem{KPZ}
There are higher order nonlinear terms of the KPZ type which will saturate
the growth at later times. However in biological cells, other processes
not included in our analysis will have intervened by this time leading to
a pinch-off of the tubule.
\end{references}
\end{document}